# Effectiveness of a time to fixate for fitness to drive evaluation in neurological patients

Nadica Miljković[1,2] · Jaka Sodnik[2]

## Abstract

We present a method to automatically calculate time to fixate (TTF) from the eye-tracker data in subjects with neurological impairment using a driving simulator. TTF presents the time interval for a person to notice the stimulus from its first occurrence. Precisely, we measured the time since the children started to cross the street until the drivers directed their look to the children. From 108 neurological patients recruited for the study, the analysis of TTF was performed in 56 patients to assess fit-, unfit-, and conditionally-fit-to-drive patients. The results showed that the proposed method based on the YOLO (you only look once) object detector is efficient for computing TTFs from the eye-tracker data. We obtained discriminative results for fit-to-drive patients by application of Tukey's honest significant difference post hoc test ($p < 0.01$), while no difference was observed between conditionally-fit and unfit-to-drive groups ($p = 0.542$). Moreover, we show that time-to-collision (TTC), initial gaze distance (IGD) from pedestrians, and speed at the hazard onset did not influence the result, while the only significant interaction is among fitness, IGD, and TTC on TTF. Obtained TTFs are also compared with the perception response times (PRT) calculated independently from eye-tracker data and YOLO. Although we reached statistically significant results that speak in favor of possible method application for assessment of fitness to drive, we provide detailed directions for future driving simulation-based evaluation and propose processing workflow to secure reliable TTF calculation and its possible application in for example psychology and neuroscience.

**Keywords** Eye tracker · Driving simulation · Neurology patients · Fitness to drive · Object detection · Time to fixate

## Abbreviations

| | |
|---|---|
| AOI | Area of interest |
| HSD | Honest significant difference |
| IDE | Integrated development environment |
| IGD | Initial gaze distance |
| lwr | Lower end point of the interval |
| max | Maximal value |
| min | Minimal value |
| MT | Movement time |
| OS | Operational system |
| PRT | Perception response time |
| PT | Processing time |
| RAM | Random access memory |
| ROI | Region of interest |
| SD | Standard deviation |
| SL | Saccade latency |
| TTC | Time-to-collision |
| TTF | Time to fixate |
| upr | Upper end point of the interval |
| YOLO | You only look once |

✉ Nadica Miljković
nadica.miljkovic@etf.bg.ac.rs

Jaka Sodnik
jaka.sodnik@fe.uni-lj.si

1 University of Belgrade – School of Electrical Engineering, Bulevar kralja Aleksandra 73, 11000 Belgrade, Serbia

2 Faculty of Electrical Engineering, University of Ljubljana, Tržaška cesta 25, 1000 Ljubljana, Slovenia

## Introduction

Evaluation of the ability to drive in neurological patients is a very challenging task for medical specialists. On the one hand, revoking a driver's license can drastically worsen a patient's quality of life, leading to social isolation, depression, and even a deterioration of physical health (Carr et al. 2019; Korner-Bitensky et al., 1994; Thompson et al., 2018). Yet, on the other hand, therapists may face potential legal liability if an individual deemed fit-to-drive is involved in an accident or jeopardizes public road safety in some other way (Korner-Bitensky et al., 1994). Consequently, decision-making on a patient's fitness to drive must be performed with exceptional caution. To make this unpleasant situation even more unpleasant, medical professionals cannot count on the reliable guidelines to evaluate fitness to drive in neurological patients as current protocols vary among countries and even among centers (Carr et al. 2019; Korner-Bitensky et al., 1994; Motnikar et al., 2020). A survey of physicians on practices around driving ability assessment in patients with nonepileptic seizures revealed that only 18% of physicians felt self-assured of their own decisions while the vast majority (93.1%) expressed the urge for evidence-based reporting guidelines (Farooq et al., 2018). This state of affairs poses an important question: Which method(s) should be used for evaluating fitness to drive by medical professionals? Evidently, well-judged decision-making consists of dedicated guidelines, as well as plausible scientific and clinical evidence (Korner-Bitensky et al., 1994).

For appropriate individual assessment, standard clinical, neurophysiological, and functional tests must be complemented with observations in a driving environment (Schanke & Sundet, 2000; Thompson et al., 2018). Indeed, proper driving evaluation for assessment of cognitive and visual functions and their complex interactions should be performed in real-life settings, as well as in response to hazards in a driving environment (Thompson et al., 2018). However, it would be way too dangerous for on-road tests to incorporate hazardous situations, especially in neurological patients (Cizman Staba et al., 2020; Edwards et al., 2003; Jurecki & Stańczyk, 2018; Motnikar et al., 2020; Olson & Sivak, 1986). Luckily, driving simulators can replace on-road tests due to their obvious safety and proven efficacy in assessing fitness to drive in neurological patients and general population (Cizman Staba et al., 2020; Frittelli et al., 2009; Motnikar et al., 2020). Furthermore, driving simulators contribute to the driving tests repeatability by providing an effective evaluation strategy for assessing a driver's behavior particularly related to risky scenarios (Ciceri et al., 2013; Fisher et al., 2007; Frittelli et al., 2009; Jurecki & Stańczyk, 2018).

We would argue that driving simulator repeatability and controllability is exceptionally important for assessing established parameters in highway engineering and road traffic safety such as perception response time (PRT). The determined norms of PRT that correspond to the reaction time in psychological literature are 2.5 s in the U.S. and 2 s in Europe. PRT is defined as the time from the first appearance of an obstacle to the initiation of braking, and presents the most important parameter associated with road accidents, especially in critical scenarios (e.g., vehicle–pedestrian collision). (Chrysler et al., 2015; Edwards et al., 2003; Green, 2000; Jurecki & Stańczyk, 2018; Olson & Sivak, 1986).

In an insightful review and meta-analysis of driving assessment of patients with Parkinson's disease (Thompson et al., 2018), it was concluded that the crash rate of collisions with pedestrians and PRT (e.g., to red lights) are, among others, commonly used outcomes. Furthermore, PRT was identified as the most promising parameter for discerning fitness to drive (Green, 2000; Motnikar et al., 2020; Schanke & Sundet, 2000), especially as a result of driving simulator complex scenery in comparison to the simple alertness tests (Cizman Staba et al., 2020; Edwards et al., 2003; Motnikar et al., 2020). Also, patients diagnosed with Alzheimer's disease, traumatic brain injury, and multiple sclerosis, have impaired driving performance as a result of, among other factors, increased PRT (Frittelli et al., 2009; Jovanović, 2021; Schanke & Sundet, 2000; Schultheis et al., 2001).

In 2019, D'Addario & Donmez proposed an interesting approach to study the effect of cognitive distraction in healthy individuals by introducing the subcomponents of the PRT (D'Addario & Donmez, 2019). By manual analysis of the eye-tracker data, D'Addario & Donmez divided the PRT into SL (saccade latency), PT (processing time), and MT (movement time) as presented in Fig. 1c. Their study emphasized that the cognitive distraction vastly influenced SLs related to the simulated road hazards. SL is defined as the time interval from the hazard appearance to the start of the first eye movement towards the hazard (Fig. 1c). In our study, we use a slightly different definition for the TTF (time to fixate) parameter, as TTF describes the time difference between the first appearance of the target object (i.e., pedestrian) in the scene to the first user's gaze on that object following procedure introduced by (Ciceri et al., 2013). Overall, PRT (Fig. 1a) incorporates detection, identification, decision, and response. Roughly, PRT can be divided into two stages: (1) perception time from the hazard onset to the onset of motion, and (2) movement time from the onset of motion to the contact with brake pedal (D'Addario & Donmez, 2019). Specifically, PRT also comprises a processing time required for the neural system to react to the perceived hazard, which starts with the TTF and continues to the movement time (for diagram simplicity, we used term "remaining processing time (PT)" in Fig. 1 to emphasize its onset). We identified two different approaches in the literature, where the first segment starts with hazard onset and finishes with either the first saccadic movement towards the hazard (D'Addario & Donmez, 2019) presented in Fig. 1c, or starts with hazard onset and ends with the first fixation on hazard (Ciceri et al., 2013) and presents time to fixate target object (Godwin et al., 2021), as shown in Fig. 1b. For computational simplicity (easier and more reliable detection of offset of TTF), we adopted the second approach presented in Fig. 1b.

Besides, our goal was to test whether the subject noticed the object rather than whether the eye movement was just directed towards or for how long it was fixed at the object of interest. To the best of our knowledge, no previous study has proposed time-to-fixate parameters in different groups

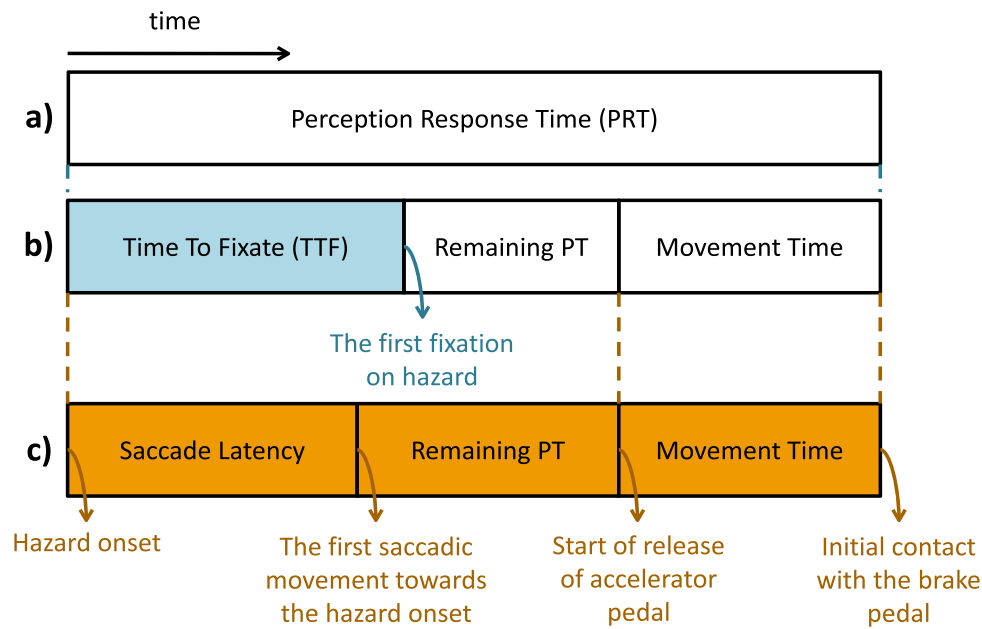

**Fig. 1** Subcomponents of the perception response time (PRT) shown in panel **a** for brake reaction time used in this paper and in (Ciceri et al., 2013, Godwin et al., 2021) comprising time to fixate (TTF), processing time (PT) marked as remaining PT as it starts with TTF, and movement time (MT) presented in panel **b** in comparison to the approach from (D'Addario & Donmez, 2019) presented in panel **c** where PRT encompasses saccade latency (SL), remaining PT, and MT

of neurological patients for assessing driving ability. We were keen on exploring the usability of this insightful TTF[1] parameter inspired by the parameters from (Ciceri et al., 2013; D'Addario & Donmez, 2019) for evaluation of fitness to drive in neurological patients. Aspects of this work have been described and tested in the master's thesis of Jovanović (2021).

**The aim of the study and contributions**

Our main aim is to present a new method for the assessment of fitness to drive of neurological patients in a simulated driving environment that incorporates eye tracking. Specifically, the method is oriented towards evaluation of the time to fixate (TTF) parameter during simulated hazard events (children crossing the road behind the bus) and the object detector is used to automate the process of TTF calculation. Our research questions are:

- Could a new method that incorporates automated object detection in conjunction with the eye tracking during driving simulation be used for calculation of TTF parameter?
- What are the main obstacles in the proposed approach and how can we enhance the future method application?
- What is the effectiveness of TTF parameter computed from the eye-tracker data in a driving simulator for discerning among three groups of neurological patients: fit-, unfit-, and conditionally-fit-to-drive (fit-to-drive 30 km around a patient's residence)?
- How do the proposed parameter relate to the standard perception response time (PRT) that is used as ground truth in this paper?

We contribute to the current body of knowledge in subsequent ways:

- We present a new method to calculate time-to-fixate parameters by adoption of the promising open-source state-of-the-art YOLO (you only look once) object detector. Software code (with sample eye-tracking video) is freely available via GitHub hosting platform with released Zenodo DOI (Miljković & Sodnik, 2023a). In addition, we provide a table with relevant parameters with open software code for statistical analysis (Miljković & Sodnik, 2023b).

---

[1] Although Godwin et al. used the same term "time to fixate" (Godwin et al., 2021), Ciceri et al. and King et al. use the term "time to first fixation" (Ciceri et al., 2013, King et al., 2019). Similar parameters describing intervals from the object appearance to the fixation were termed *saccade latency* and *perception time* (D'Addario & Donmez, 2019). For details on the applied definition for time to fixate (TTF), please see the main text and for other variable names such as saccadic reaction time, we recommend a remarkable guideline in Holmqvist et al., (2022). The presented explanations of the calculated parameter comply with the good practices and guiding standards in avoiding research pitfalls proposed in Godwin et al., (2021) with the main aim to apply systematic manner in defining eye-tracking metrics by avoiding the use of different terms to describe already-established parameters (Godfroid & Hui, 2020; King et al., 2019).

- We give a workflow for automated detection of the time-to-fixate parameter from the eye-tracker video in pedestrian hazard simulations. We also explain encounters as well as potential ups and downs of the proposed approach, mainly in regard to the exacerbated noise in the eye-tracker videos and to the YOLO accuracy as we identify both of these as main bottlenecks in the future applicability of the proposed method and for study reproducibility.
- We test whether time to fixate is a valuable parameter for discerning among three groups of neurological patients and we address its possible future application and usability in assessment procedures involving driving simulators or similar complex scenarios.

Altogether, we advocate that the method presented in this paper and tested for the eye-tracker videos in a heterogeneous group of neurological patients could be employed in studies of cognitive attention in related research fields such as experimental and cognitive psychology, architecture, neurosciences, etc. At the same time, we provide methodological concerns for future method implementation, especially considering that the present study is of retrospective nature i.e., a posteriori study. We also aim to comply as much as possible with the good scientific practices for ensuring valid eye-tracking research outcomes (Godfroid & Hui, 2020; Godwin et al., 2021; King et al., 2019; Orquin & Holmqvist, 2018).

## Materials and methods

The eye-tracker data were recorded in 108 neurological patients. All patients participated in a standard procedure for license revalidation at the University Rehabilitation Institute Soča in the Republic of Slovenia. At the facility, patients underwent clinical, neurophysiological, functional, and on-road evaluation where the following scores were assigned to each patient: fit-, conditionally-fit- (fit-to-drive 30 km around patient's residence), and unfit-to-drive. After initial recruitment, due to the unavailability of the data, assessment was performed on 91 subjects (24 females and 67 males) with ages ranging from 18 to 89 years (49.88 mean and standard deviation of 17.15 years). Overall, 33 subjects were diagnosed with traumatic brain injury, 35 with non-traumatic acquired brain injury, and 21 had neurodegenerative diseases (14 with multiple sclerosis, four with Parkinson's disease, and three with other diseases). The remaining two patients were diagnosed with Guillain–Barré syndrome and epilepsy. The study included subsidiary testing of patients' ability to drive in a motion-based driving simulator produced by Nervtech Ltd. (Ljubljana, Slovenia). All patients signed informed consent forms in accordance with the Code of Ethics of the World Medical Association (Declaration of Helsinki) and the study was approved by the Medical Ethics Committee at the University Rehabilitation Institute Soča. (Motnikar et al., 2020).

The videos available from the Tobii Pro Glasses 2 (Tobii Llc., Stockholm, Sweden) with sampling frequency of 50 Hz were used in order to automatically calculate TTF parameters and to manually determine whether an actual collision occurred. We used only one scenario for TTF calculation, as it incorporates common pedestrian collision. The scenery comprised a rural area where participants were expected to brake abruptly in response to two children running out behind a stopped bus (Fig. 2) (Cizman Staba et al., 2020). Other scenarios lacked the essential "surprise factor" that the selected scenario had with the children running behind the bus, so TTF was calculated only for the selected collision.

The scene comprised a vehicle moving in a straight direction towards the bus station at a speed that should be less than or equal to 50 km/h. As the vehicle approaches, children that were hidden behind the parked bus on the right side of the road start to run across the street. The driver is expected to perceive the danger and decelerate. A road consists of two lanes in opposite directions and without any traffic lights. On the right side of the road, two traffic signs indicate "crosswalk ahead" and "children crossing road" signs. Weather conditions were suitable for driving (sunny weather without rain/snow/fog, see Fig. 2).

We would like to emphasize here that this is a retrospective study and that we were not able to design a protocol and to control the driving simulation. Strictly speaking, the bus scenario is not created with the aim of TTF assessment, as the children's appearance was determined by the internal simulator time and it was independent of the driver's performance. Luckily, in the majority of cases, drivers behaved as instructed in a similar way, probably as they were aware of the testing procedure for their fitness evaluation. Other scenes were closely inspected by the authors and although two scenes incorporated abrupt appearance of pedestrians, they were excluded from the analysis, as the pedestrians in the majority of cases were too close to the vehicle and as corresponding pedestrians were programmed never to cross the road, so the actual hazard did not take place. Besides, additionally inspected scenes were not from the same rural area, but from a city area. Therefore, a fair comparison cannot be made, as such dynamic environments have different visual properties. Similar statements can be made regarding the comparison of static images with varying visual properties in Godwin et al., (2021) and Orquin & Holmqvist, (2018).

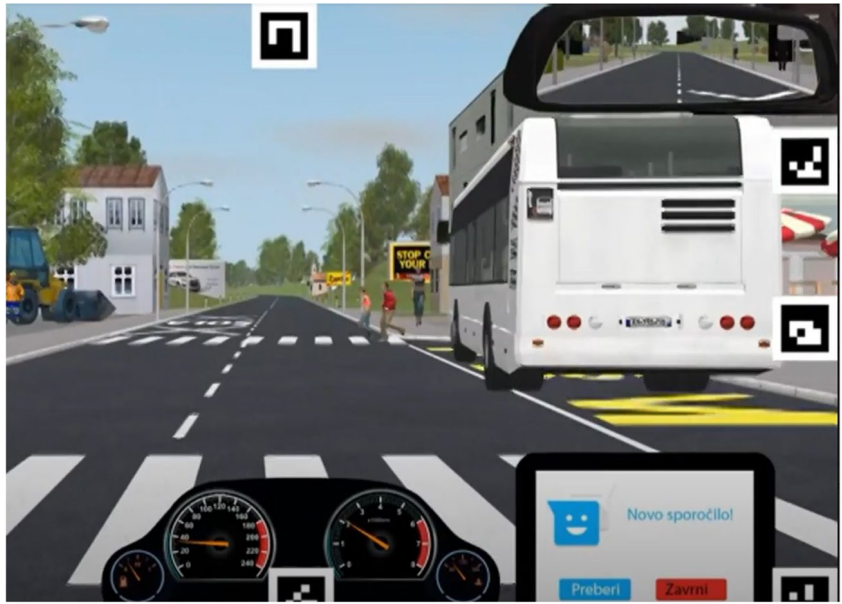

**Fig. 2** Sample snapshots from the Nervtech simulator for the selected collision scene with children running behind a parked bus

## Video analysis and feature extraction

All processing steps were performed in Python v3.8 (Python Software Foundation, Wilmington, DE, USA). We used Visual Studio Code (Microsoft, Redmond, WA, USA) as an IDE (integrated development environment) for video analysis on computer with Linux Ubuntu 20.04 OS (operational system) with 8 GB RAM (random access memory) and Intel Pentium 3556U with 1.7 GHz. Apart from standard libraries for data processing, such as numpy (Oliphant, 2006) and scipy (Virtanen et al., 2020), a specialized OpenCV module (Bradski & Kaehler, 2008) for video editing and analysis was used. YOLO v5s object identifier (Jocher, 2020; Redmon & Farhadi, 2018; Redmon et al., 2016) was used for object detection due to its proven outstanding possibilities to annotate objects and their respective probabilities with relatively high accuracy. Since its introduction in 2016 by Redmon et al. (Redmon et al., 2016), YOLO received much attention in the scientific and technical communities. In addition to the welcoming advantage of YOLO open-source license, the main YOLO supremacy is its speed (from 45 up to the 155 frames per second for real-time employment). Compared to other detection systems, YOLO makes more localization errors, but is less likely to predict false positives on backgrounds, which is extremely important in traffic situations (Redmon et al., 2016). In addition, it is also known that YOLO may have larger error rates (> 30%) as a consequence of smaller object size, which was exactly our case when pedestrians entered the scene (Redmon et al., 2016). Though some of the previous flaws were corrected in YOLO version v3 (Redmon & Farhadi, 2018), this challenge remains in v5 as well (Al Amin and Arby, 2022) that is used in our study.

To test the YOLO accuracy in pedestrian detection and to ensure reliable and correct TTF calculation, manual checks of already-trained YOLO classifiers were conducted and possible TTF errors due to inappropriate YOLO localization were manually corrected. This was performed by manual checks of the corresponding scenes for the exact frame when the gaze was captured within the object boundaries that were mistakenly localized by YOLO to either other close objects on the scene or were absent completely (failed labeling or failure to detect the object). The problem of detecting small objects in a scene is very well known in computer vision, especially in aerial image objects (Liu et al., 2021). In this study, we had to introduce manual checks and corrections in ~ 36% (20 out of the selected 56 recordings) due to the YOLO errors. Future efforts towards automated procedures may be focused on the customization of openly available YOLO detector, but here we aimed to present YOLO perspective for eye-tracker data analysis and TTF calculation as is.

The selected scenario was segmented from the eye-tracker video with OpenCV tools and then each video segment comprising collision with pedestrians was converted to image sequences i.e., frames. Further, each frame was fed to the YOLO object detector to localize ROI (region of interest)[2]

---

[2] We use the ROI abbreviation for region of interest being a common term in object-detection studies. However, AOI (area of interest) is also routinely used in eye-tracker-related research (Holmqvist et al., 2022).

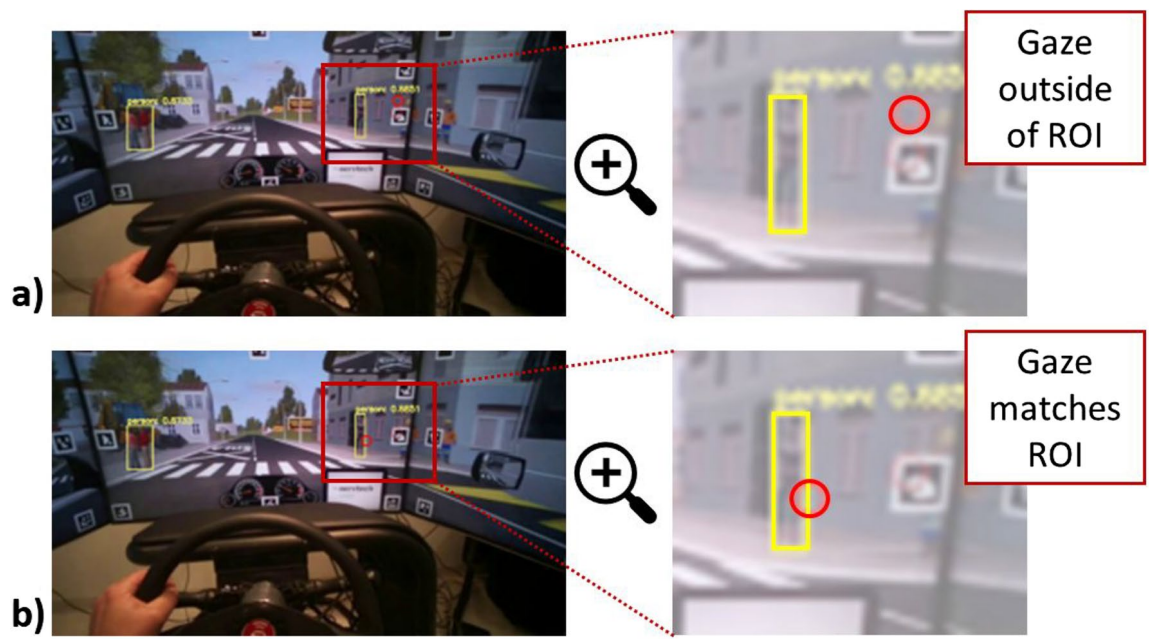

**Fig. 3** Sample snapshots from the YOLO detector with the eye-tracker gaze for pedestrian detection with enlarged details in a random simulator scene. **a)** Snapshot of subject's gaze (*red circle*) outside of region of interest (ROI) marked by the *yellow rectangle* and determined by the YOLO detector as "person" b. Another snapshot of the subject's gaze that matches ROI with the pedestrian on the right pavement. **b)** The process used to mark the event when subject notices the object to calculate time to fixate (TTF) parameter

i.e., persons in an image. Sample output snapshots from the YOLO detector are given in Fig. 3. YOLO output is a rectangle with annotated ROI in the relevant image. For illustration of YOLO output and its cross section with the eye-tracker gaze (red circle in Fig. 3), we used a sample scene in Fig. 3 that does not correspond to the selected hazard scene with the children running behind the bus in Fig. 2.

To calculate TTF, we subtracted the timestamps *t2* and *t1*, where *t2* presented the timestamp when driver noticed children. Namely, *t2* corresponds to the timestamp when the driver's gaze falls within the detection frame in a manner shown in Fig. 3b). Precisely, we hypothesized that driver notices the pedestrian at the moment when the driver's gaze foveates ROI comprising a pedestrian, as presented in Fig. 3b. Timestamp *t2* corresponds to the time prior to the driver's response (deceleration and braking), as shown in Fig. 1b (please, see the marked moment of the first fixation on hazard). We determined *t1* as the time when children began to run across the road. Precisely, *t1* presents the time when the YOLO detector identified children appearance in the video frame. The complete video analysis workflow block diagram for TTF computation is presented in Fig. 4 and freely available in Miljković and Sodnik (2023a).

Overall, 35 eye-tracker videos were excluded from the study due to the following conditions that prevented detection of the time to fixate parameter: tilted glasses (eight patients were excluded), missing data (11), due to the anticipatory pedestrian perception (two), altered glasses positioning (four), as eye gaze could not be determined either due to the gaze information loss or frozen gaze for the entire segment (seven), due to the pauses in recording that prevented further analysis of the particular scene (two), and in one case the participant almost ran over the kids, as they did not look at pedestrians at all, so the TTF could not be calculated (one). The remaining valid 56 recordings were analyzed, and they consisted of 20 fit-, 17 conditionally-fit, and 19 unfit-to-drive patients, presenting a relatively balanced group distribution for further analysis. Previous results on Tobii device data quality by Niehorster et al., (2020) showed that the signal was relatively stable, but not robust, while subjects spoke, made facial expressions, and moved the eye-tracker device. Our participants were seated in a controlled environment, but they were not constrained from commenting or making facial expressions and therefore the data quality may be partly affected. Although some manufacturers provide corrective procedures and post-analysis to compensate for noise, eye-tracker devices can still move in respect to the subject's head causing the "slippage" that can drastically deteriorate the signal quality regardless of the type of the eye-tracker device (Niehorster et al., 2020). The Nervtech driving simulator is a motion-based device and the incorporated haptic feedback may have added to the "slippage" of the eye-tracking setup in our case. Therefore, we were especially cautious in interpreting video signals that interfered with "slippage" by introducing manual verification procedures and "slippage" influence on the gaze

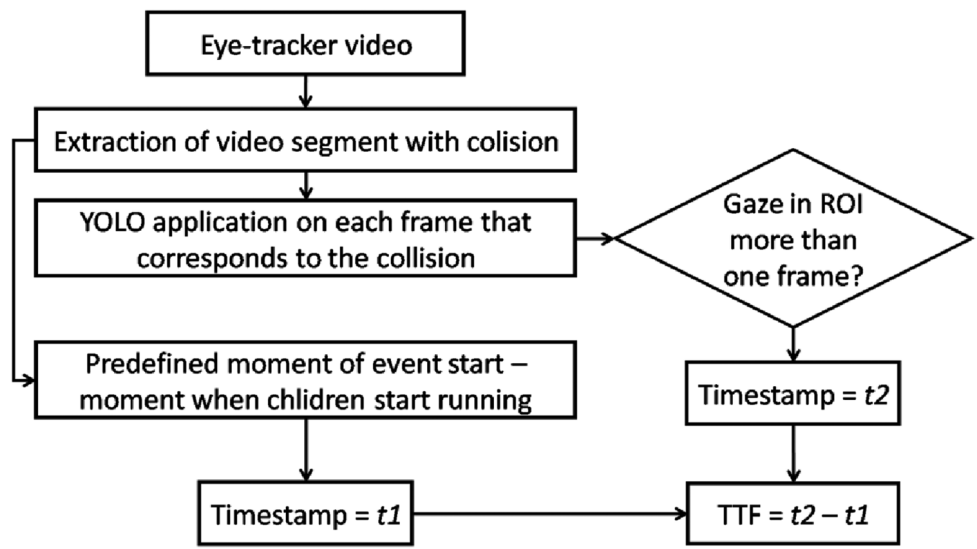

**Fig. 4** Block diagram of automated video processing workflow with YOLO (you only look once) object detector for calculation of TTF (time to fixate) parameter. *ROI* stands for region of interest. For more details, see main text

position prior to video analysis. Moreover, Tobii eye trackers have already demonstrated the data loss in the altered gaze direction (Niehorster et al., 2020), which is also confirmed in our study, and we had to exclude data from four participants due to the eye-tracker positioning that affected gaze direction in respect to the device.

As proposed in (D'Addario & Donmez, 2019), we also extracted additional features to test their influence on TTF. Auxiliary features are: (1) the participants' speed at hazard onset, (2) initial gaze distance (IGD), and (3) time-to-collision (TTC). The children's position at the collision onset was estimated using the YOLO object detector as the center of the ROI rectangle with ($X$, $Y$) coordinates in horizontal and vertical directions in pixels. The overall frame resolution of the eye-tracker video was $960 \times 540$ pixels. Then, the distance between the children's position and initial gaze coordinates was determined as Euclidian distance in pixels and rounded to the nearest integer value. The obtained IGD value was then transformed by the common pixel-to-angle transformation factor based on the Tobii Pro Glasses 2 User's Manual (Tobii Pro, 2020). The TTC parameter was calculated as the ratio between the distance of the ego vehicle to the pedestrians and its speed. The distance between the ego vehicle and pedestrians was recorded automatically by the simulator. We added speed, IGD, and TTC as covariates to test their impact on statistical significance of TTF parameter for differentiation of three groups of patients.

Since this is a retrospective study, we had no means to control the protocol or create specific driving scenarios for isolating and measuring the TTF parameter. We therefore also included PRT as a ground truth parameter to observe the relationship between TTF and PRT, since our previous research preformed on the same dataset already demonstrated a correlation of PRT time collected in the driving simulator and results of standardized neurophysiological assessment tests (Cizman Staba et al., 2020; Motnikar et al., 2020). Moreover, reaction times are in general commonly analyzed by researchers (Godwin et al., 2021). In our driving scenario, the stop sign was introduced three times during the driving session (large stop mark appeared over the central screen and subjects were instructed to stop the vehicle). PRTs were extracted from the simulator log files as the times from predefined trigger (stop sign appearance) until the drivers pressed the brake with the force of at least 75 N. Moreover, PRTs are determined for the participants in whom we calculated TTF.

### Statistical analysis

For statistical analysis, we used R programming language v4.1.2 (Team R. C., 2021) in the RStudio environment (RStudio, Inc., Boston, USA) with dplyr (Wickham et al., 2015), ggplot2 (Wickham 2016), and ggpubr (Kassambara 2023) packages for analysis and visualizations. As TTFs did not conform to the assumption of normal distribution, we used Welch's ANOVA for our analysis following the approach applied also by Motnikar et al. on the same patient sample (Motnikar et al., 2020). We also used Tukey's honest significant difference (HSD) post hoc test with a confidence level of 0.95 to create a set of confidence intervals on the means differences. Similar to Motnikar et al., (2020), TTFs deviating for more than three interquartile ranges from the third quartile were considered outliers and were removed from the analysis. Moreover, valid TTFs were considered

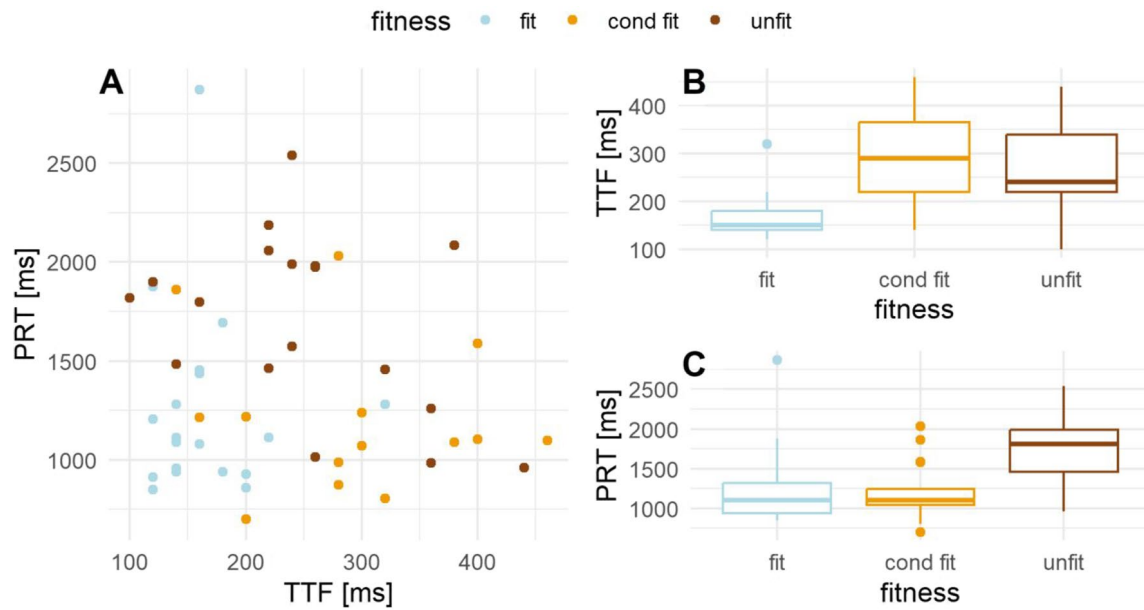

**Fig. 5** Scatter plot of TTF (time to fixate) and PRT (perception response time) parameters are presented on panel **A**, while separate boxplots for TTF and PRT are presented in panels **B** and **C**, respectively, for the three groups of patients. Fit, unfit, and cond fit stand for fit-, unfit-, and conditionally-fit-to-drive groups of patients, respectively

to be below 500 ms, as longer reaction times were previously considered as a "miss" – drivers not paying attention to the road and reacting too late or not at all (Cizman Staba et al., 2020; Edwards et al., 2003). Altogether, three observations were lost. For the obtained PRTs, we adopted the same identification of outliers with interquartile range, and we considered them valid only in the range from 0.5 s and 4 s as proposed in (Cizman Staba et al., 2020) prior to averaging three PRT repetitions. In one unfit patient, PRTs were missing for all three repetitions. In cases when one or two PRTs were missing, we were still able either to calculate the mean or to use the remaining PRT. As the remaining PRTs conformed to the normal distribution, we used the *F*-test for the equality of means in one-way ANOVA. As for TTF, we used Tukey's honest significant difference (HSD) post hoc test with confidence level of 0.95 on PRT parameters.

Additionally, a general linear model (multi-factor ANOVA) was also performed to see if fitness to drive, speed, IGD, and TTC explain variation in patients' TTF with rigor significance level of 0.001 to compensate for non-homogeneity of TTF across groups of neurological patients. We used Shapiro–Wilk test to check the normality of model residuals. Also, Pearson's product moment cross-correlation coefficients between TTF and speed, IGD, TTC, and PRT were calculated to test whether speed, IGD, TTC, and PRT are linearly related to TTF. Tables with all relevant parameters and R code are available in Miljković & Sodnik, (2023b). Summary statistics for all parameters are reported in the Results section.

We assessed measurement uncertainty type B for TTF calculation, as well as combined measurement uncertainty according to the Guide to the Expression of Uncertainty in Measurement (BIPM et al., 1993) from the temporal resolution determined by the sampling frequency of the eye-tracker video under the assumption of the uniform distribution.

## Results

To get visual insight into the differences of TTF and PRT parameters, as well as an understanding of their relationship, we provide scatter plots of PRT and TTF in Fig. 5a and boxplots of TTF and PRT in Fig. 5b and c, respectively. In Fig. 5, separate colors are used to discern among three groups of neurological patients: light blue for fit-, brown for unfit-, and orange for conditionally-fit-to-drive patients.

The results of Tukey's HSD post hoc test for TTFs and PRTs with confidence level of 0.95 are presented in Table 1 and Table 2, respectively. Welch's ANOVA for mean comparisons among groups resulted in $F = 15.935$ and $p = 0.000$ for TTF, while in $F = 7.038$ and $p = 0.002$ for PRT indicating the differences among groups as visually revealed in Fig. 5b, c.

Tukey's honest significant difference post hoc test for TTF parameter from Table 1 revealed that the fit patient group was significantly different from unfit ($p = 0.001$) and conditionally-fit-to-drive ($p = 0.000$) groups. However, no significant differences were attained between unfit and conditionally-fit-to-drive patient groups ($p = 0.542$). A clear distinction of interval for differences of the observed means for statistically significant results is shown in Table 1, and corresponds to the TTF box plots in Fig. 5b. Interestingly, PRT failed to detect differences between conditionally-fit-to-drive and fit-to-drive ($p = 0.983$), while the unfit-to-drive group showed significant differences from the other two groups (Table 2). This is not just in line with PRT boxplot in Fig. 5c, but also reveals a possible relationship between TTF and PRT corresponding to the scatter plot in Fig. 5a.

**Table 1** Results of Tukey's honest significant difference post hoc test with confidence level set at 0.95 for TTF (time to fixate) parameter. Differences in the observed means, lwr (lower end point of the interval), and upr (upper end point of the interval) are presented as integers. Statistically significant results are highlighted in bold. We report 7 decimals for adjusted $p$ for more precise comparison

| TTF | Difference [ms] | Lwr [ms] | Upr [ms] | Adjusted $p$ |
| --- | --- | --- | --- | --- |
| Fit vs. Conditionally-fit | – 130 | – 199 | – 61 | **0.0001074** |
| Unfit vs. Conditionally-fit | – 31 | – 101 | 39 | 0.5421314 |
| Unfit vs. Fit | 99 | 35 | 163 | **0.0012709** |

To test the reliability of the performed statistical analysis, we repeated statistical tests with included TTF outliers and concluded that statistically significant discrimination among groups did not change ($p = 0.000$ and $p = 0.037$ for significant differences, while non-significant had $p = 0.268$).

Summary statistics (parameters range containing minimal and maximal values, mean values with standard deviation, and median value) for the TTFs, speeds, IGDs, TTCs, and PRTs for three groups of patients are presented in Table 3. By just looking at median values (we focus on them as the most relevant in cases that parameters do not conform to the normal distribution), it is obvious that the highest relative changes are reported in Table 3 for TTF and PRT, while other parameters remained relatively similar. Also, we should emphasize the relatively large standard deviations for all parameters being the smallest for the PRTs.

For three recordings, participants were not focused on driving activity and consequently the TTFs were relatively large, so we excluded them as outliers according to the criterion described in the previous section (720, 1420, and 1060 ms). Therefore, we did not incorporate outliers in summaries in Table 3. All excluded TTF outliers belonged to patients from the conditionally-fit-to-drive patient group.

Measurement uncertainty type B for TTF is 5.77 ms for temporal resolution of 20 ms determined from the eye-tracker video sampling frequency (50 Hz). As standard deviations of TTFs were much larger TTFs ($> 47$ ms from Table 2) than the measurement uncertainty type B, we discarded measurement uncertainty type B from further analysis.

A general linear model was run to also observe fitness to drive, speed, IGD and TTC on TTF. While controlling for other parameters, speed ($p = 0.224$, $F = 1.588$), IGD ($p = 0.875$, $F = 0.025$), and TTC ($p = 0.015$, $F = 7.296$) do not significantly predict patient's TTF, except for the fitness to drive ($p = 0.000$, $F = 25.079$). Shapiro–Wilk normality test confirmed our assumption that residuals follow normal distribution ($p = 0.601$). We find significant interaction among fitness, IGD, and TTC ($p = 0.001$, $F = 10.697$) on TTF. Cross-correlation coefficients revealed no significant correlation of speed ($-.105$, $p = 0.503$), IGD ($-0.004$, $p = 0.979$), TTC ($0.290$, $p = 0.059$), and PRT ($-0.085$, $p = 0.55$) with TTF.

Crash rates showed that only one subject dimmed conditionally fit-to-drive crashed into the pedestrians (with TTC = 1.7 s and speed 47 km/h). The results for this subject were not included in the final analysis due to the eye-tracker tilt. Overall, 6 out of the analyzed 56 subjects (10.7%) almost crashed into simulated pedestrians *i.e.*, avoided crash, but had to abruptly stop in immediate proximity to children. Two of them were unfit-, two conditionally fit-, and two even fit-to-drive. Four of these patients had very low TTC of either 0.9 or 0.8 s, while one unfit-to-drive patient had TTC of 4.7 s, and for one we could not determine TTC due to the lost data.

**Table 2** Results of Tukey's honest significant difference post hoc test with confidence level set at 0.95 for PRT (perception response time). Differences in the observed means, lwr (lower end point of the interval), and upr (upper end point of the interval) are presented as integers. Statistically significant results are highlighted in bold. We report 7 decimals for adjusted $p$ for more precise comparison. Values are rounded to integers

| PRT | Difference [ms] | Lwr [ms] | Upr [ms] | Adjusted $p$ |
| --- | --- | --- | --- | --- |
| Conditionally-fit vs. Fit | – 25 | -370 | 320 | 0.9831483 |
| Unfit vs. Fit | 455 | 115 | 794 | **0.0059359** |
| Unfit *vs.* Conditionally fit | 480 | 127 | 833 | **0.0052338** |

**Table 3** Summary statistics of parameters for the three groups of neurological patients (fit, conditionally-fit, and unfit-to-drive). TTF is time to fixate, SD stands for standard deviation, IGD for initial gaze distance, TTC to time-to-collision, and PRT for perception response time. For calculating SDs, we used Bessel's correction. See text for more details

| Parameters | Calculated values | Fit | Conditionally fit | Unfit |
|---|---|---|---|---|
| TTF [ms] | Range | 120–320 | 140–460 | 100–440 |
| | Mean (SD) | 163 (47) | 293 (95) | 262 (99) |
| | Median | 150 | 290 | 240 |
| Speed [km/h] | Range | 7.4–46.9 | 5.5–34.8 | 3.3–41.1 |
| | Mean (SD) | 19.5 (10.5) | 19.1 (11.4) | 20.3 (10.9) |
| | Median | 15 | 15.3 | 19.6 |
| IGD [°] | Range | 0.82–21.99 | 0.54–8.36 | 0.91–23.71 |
| | Mean (SD) | 6.67 (6.35) | 3.90 (2.20) | 4.99 (4.76) |
| | Median | 3.86 | 3.81 | 3.13 |
| TTC [s] | Range | 0.7–10.1 | 0.5–9.6 | 0.8–19.6 |
| | Mean (SD) | 3.3 (2.6) | 3.9 (3.5) | 4.2 (4.7) |
| | Median | 2.9 | 2.7 | 2.4 |
| Average prts [ms] | Range | 850–2871 | 700–2031 | 960–2540 |
| | Mean (SD) | 1241 (475) | 1215.98 (358) | 1696 (449) |
| | Median | 1101 | 1104 | 1808 |

## Discussion

The time-to-fixate parameter proved its usability in discerning fit-to-drive patient group and at the same time did not capture the difference between conditionally-fit- and unfit-to-drive patients. Although unexpected, this is a rather interesting outcome, as previous study on the same group of patients failed to find differences between fit- and conditionally-fit-to-drive groups of neurological patients (Motnikar et al., 2020), which is also confirmed in the same PRT parameter in the presented repeated analysis. We would argue that these two findings are complementary, as conditionally-fit-to-drive patients are indeed at the borderline between fit- and unfit-to-drive groups. Exploration of possible application of machine learning algorithms for investigating more in-depth this multiplex interaction in larger and more diverse samples would be an attractive future approach, especially if we take into account the nonlinear class separability in the scatter plot in Fig. 5a. The fit group is characterized by lowest TTF and PRT and the conditionally-fit-to-drive group with higher TTF and lower PRT. The unfit-to-drive group on the other hand revealed mixed results, as the majority had low TTF and high PRT, while some participants had both values that were high. This can partly be explained by the compensatory mechanism in conditionally-fit patients that enables them to perform well with the faster motor reactions in demanding situations, while unfit-to-drive patients have consistently prolonged PRTs and are therefore unable to perform the task. In what follows, we thoroughly discuss obtained results and their possible implications with awareness that TTF although showed statistically significant discernment among groups of neurological patients may be prone to other effects. We strived to report all factors that could possibly affect reproducibility of the proposed method aiming to address contemporary scientific challenges such as questionable measurement practice (Godwin et al., 2021; Flake & Fried, 2020), variable data quality (Hessel & Hoodge, 2019), and reproducibility crisis (Open Science Collaboration, 2015).

### Prospects of YOLO-based eye-tracker video analysis

For the exact localization, YOLO can struggle to reach high accuracy, with error rates of up to 34.5%, especially in cases of smaller objects in a scene (Redmon et al., 2016). This error rate is in line with our results of ~ 36% of inaccurate ROI recognition. Still, we argue that the level of the fidelity of the driving simulation may also play a significant role in pedestrian detection by the YOLO identifier (Redmon et al., 2016). YOLO is a general-purpose detector, and for pedestrian detection, some improvements of initially pretrained detectors have already been proposed. For example, Valiati & Menotti introduced weak semantic segmentation in the learning phase to enhance YOLO performance for pedestrian detection (Valiati & Menotti, 2019). More examples include tiny-YOLO improvements proposed by Yi et al. (2019) and improvement of network layers by Lan et al. (2018) for pedestrian detection. Additionally, one of the main challenges in computer vision is the detection of small objects in a scene. Commonly, the term "small" is used to indicate small in the complete scene, especially in aerial computer vision. One of the promising enhancements of object detectors incorporates artificial augmentation by replication of small objects in images (Kisantal et al., 2019) while an outstanding review of challenges for small objects detection is given in (Liu et al., 2021). Hence, future approaches may include adaptations to the existing YOLO versions for more accurate pedestrian detection, especially in the simulation and in cases when the object size is relatively small. Although our optimistically proposed approach of applying non-extended YOLO detector for automated TTF calculation (Fig. 4) had a relatively high error rate, the proposed semi-automatic detection is at least one step ahead and less time-consuming of existing hand-operated procedures (D'Addario & Donmez, 2019).

Conclusively, we believe that enlargement of the input parameters of the pretrained YOLO would be a reasonable way to improve its performance for detection of simulated pedestrians. Therefore, reliable open detection benchmark data for driving simulators are required. On one side, this is an especially exciting direction, as novel YOLO versions are focused on automated vehicles with appropriate accuracy rate for small object detection (Mahaur & Mishra, 2023), and thus can incorporate complex human–machine interactions by introducing parameters such as TTF. On the other side, real-time assessment in simulators can speed up the decision-making on fitness to drive in neurological patients with immediate evaluation in complex environments. Until then, the proposed YOLO-based automated approach (Fig. 3) with manual inspection presents a relatively fast and reliable technique to determine TTF.

### Eye tracker video analysis: Ups and downs

Although 108 patients initially agreed to participate in the study and 91 entered eye-tracker video analysis, we had very high data loss due to the unreliability and unsteadiness of the eye-tracker glasses. In a previous study on the same sample (Motnikar et al., 2020), reaction times in response to the on-screen instruction to come to a full stop were calculated for 54 subjects due to the "technical error". Here, we were able to calculate TTFs in 56 patients being just two more than reported in (Motnikar et al., 2020). Apparently, the eye-tracker data quality was assessed in a similar manner. Moreover, this sample size proved sufficient to achieve comparable statistical power with the previous study on the same sample.

Although measurement uncertainty type B is relatively small, as a result of resolution of 20 ms, our analysis may have hindered the gaze accuracy by other effects such as varying distances already reported by the manufacturer. In the course of writing this article, an empirically based minimal reporting guideline for the eye-tracker was published in Holmqvist et al., (2022). We performed a post-assessment of our method to verify whether our procedure conforms to the noteworthy instructions for eye-tracking studies proposed in Holmqvist et al., (2022). Subsequently, we stress that overall recording environment comprising setup and geometry, measurement space and monitor size, distance between participant and the eye tracker were all kept as constant as possible, although we could not control for every single possible effect. Future procedures might consider either appropriate corrections of the recorded signal or automatic opt for eliminating such trials in order to preserve data quality that can vastly influence the research results (Holmqvist et al., 2012; Holmqvist et al., 2022). Our results show that despite all potential influences, the TTF parameter was able to discern fit-to-drive group of patients. Nonetheless, there is still room for method improvements and future directions that should be focused towards obtaining reliable TTF parameters. We suggest the following directions, based on our own experiences and on the relevant literature: (1) the choice of the eye tracker, with the focus on higher resolution i.e., sampling frequency is of tremendous importance for the correct fixation/gaze/saccade/glissade detection (Leube & Rifai, 2017; Nyström & Holmqvist, 2010) and (2) reliable measurement that goes hand in hand with the selection of appropriate processing methods (Nyström & Holmqvist, 2010) that can be applied for more reliable detection of eye-tracking events. Tests in a laboratory setting with, for example subjects instructed to limit their movements, is unfortunately not an option, as natural subject behavior with the "surprise" element (e.g., hazard scene) is desirable for appropriate driving simulation assessment (Chrysler et al., 2015). However, adaptive movement artifact elimination with techniques commonly used in electrophysiology (Raya & Sison, 2002) could be explored for application also in eye-tracking data. For example, additional accelerometer measurements have already proven usable in measurements with electrooculography googles to compensate for movement artifacts (Bulling et al., 2009). We ensured that glasses were calibrated at the beginning of the measurement to compensate for scene instability in the eye-tracker videos, but this may be (and should be) further improved by adopting mentioned strategies.

To check whether fitness of the patients is related with manual corrections that together with small object size (as explained in the section Prospects of YOLO-based eye tracker video analysis) could influence YOLO analysis, we performed additional checks of how many manual checks were in each group of subjects (the code is available in Miljković & Sodnik, (2023b). Results were inconclusive, as unfit-to-drive patients had higher data ratios without manual corrections (68.4% of all unfit-to-drive patients and 23.2% of all patients that were analyzed), indicating that their eye-tracker data are the most dependable. Fit patients showed slightly lower results (60.0% and in overall sample 21.4%), while conditionally-fit-to-drive patients had the lowest ratio without manual corrections (47.1% and in overall sample 14.3%). For excluded patients, we could not make any conclusions based on the fitness relation with saccade accuracy. However, in our experience, unfit-to-drive patients tend to adopt very unnatural behaviors, such as driving too slowly or too many fixations during the simulation probably in an attempt to compensate for altered neural function, which is already confirmed in the elderly population (Green, 2000), so the data quality assessment and analysis of other eye-tracking parameters may be a perspective for such patients. Regardless of subjects' individual driving strategies, gaze precision for TTF parameter calculation is of tremendous importance, as accidental gazes could vastly influence our results, especially the determination of whether the gaze is within the ROI. We did go through the videos to check for such cases, but we did not apply any systematic approach. Future improvements of the proposed method may include more thorough analysis of both saccades and fixations, as the pedestrian fixation should be preceded by a saccade unless the subject's initial gaze is on the place where the pedestrian appeared. This also gives us an insight into what is the fastest possible TTF, as the time stamp *t2*, used to mark the end of the TTF parameter, should be preceded by a saccade. Then, TTF should be at least 120 ms, which is the saccade duration reported by the ISO 15007 standard. Only one TTF was lower than 120 ms, but in the end we decided to keep this sample throughout the analysis. Also, future studies could improve gaze detection by manipulation of the driver's attention prior to triggering the pedestrian's appearance.

### Is time to fixate a useful parameter for evaluating fitness to drive?

Direct comparison of TTFs could be made only with the so-called SL presented in (D'Addario & Donmez, 2019) and for time to first fixation in (Ciceri et al., 2013), both assessed only in healthy participants. Namely, SLs for pedestrian hazards were 0.34 s (mean), 0.37 s (median), 0.00 s (min), 0.58 s (max), and 0.17 s (SD), which are partly in line with our results of 0.163 s (mean), 0.150 s (median), 0.120 s (min), 0.320 s (max), and 0.470 s (SD) for fit-to-drive patients. SLs in healthy subjects are, to our surprise, more in accordance with TTFs in conditionally-fit-to-drive patients. However, these comparisons should be taken with a grain of salt. In other words, PRT cannot be explored independently of the road collision situation, as even tiny variations may have a significant effect on collision management (Ciceri et al., 2013; Jurecki & Stańczyk, 2018). We would stress that the same principle applies for the TTF parameter as existing evidence for time to first fixation reveals direct relation to the testing conditions (Ciceri et al., 2013). Mean times to first fixation ranged from 0.06 to 0.53 s with SDs from 0.14 to 0.42 s being to some extent in line with our results. The test variations that could cause shorter TTFs in our study may be related to the fact that two children pedestrians were running across the street behind the bus. Consequently, the test could contribute to the urgency of the risk avoidance and visual detection. This is in accordance with previous studies where faster-moving pedestrians presented a more challenging response scenario by reducing TTC (Chrysler et al., 2015). Although driving simulators present an excellent tool to evaluate driver's behavior in a real-life complex environment, the presence of the investigators and somewhat unusual conditions may have generated more caution in subjects (Olson & Sivak, 1986), which is in our case especially visible for speed before the collision which was relatively low (Table 3). Compensate mechanism could have taken place in this slow drive as well, as older people tend to respond slowly, in general, and to counterbalance for reduced cognitive skills by lower speeds on the road (Green, 2000).

In Cizman Staba et al., (2020), reaction times were considered valid in a range of 0.5–4 s, as shorter and longer reaction times were considered "cheat" and "miss", respectively. Similarly, in Edwards et al., (2003), the PRT could not be calculated in some cases due to the lack of response or due to the advanced reaction. By following and adapting this approach to TTF, we excluded TTFs that were larger than 500 ms, as a typical glance is between 500 ms and 3 s, as stated in the ISO 15007 standard. Despite this rejection, our main conclusion related to fit-to-drive discrimination did not change.

TTF is a subdivision of a response or PRT (Fig. 1), which can be influenced by a variety of factors and diseases. For example, mean PRT to vehicle block was slow after alcohol (2.21 s) and fexofenadine (1.95 s) consumption (Weiler et al., 2000). Also, some mixed findings were reported on driver's age on PRT, but the majority of studies agree that increased age is related to increased PRT (Broen & Chiang, 1996; Edwards et al., 2003; Lerner, 1993; Olson & Sivak, 1986). Reported PRTs for pedestrian's sudden appearance in the intersections for the young age (19–23 years) group was 0.97 ± 0.46 s and for the older group (65–83) it was 1.44 ± 0.45 s (Edwards et al., 2003). These PRTs are higher than the TTFs presented in this study, which is expected. In subjects with Alzheimer's disease, mean visual reaction time was 511.00 ± 63.20 ms, while healthy controls had 390.00 ± 29.50 ms and age-matched neurological normal controls had 384.00 ± 31.80-ms reaction time (Frittelli et al., 2009). This suggests that fit-to-drive patients could correspond to more healthy-like "normal controls" in (Frittelli et al., 2009), indicating possible differentiation among patient groups by reaction times and reasoning for our findings.

No significant mean effect was observed between conditionally-fit- and unfit-to-drive patients (Table 2). It is stated that patients tend to focus visual attention to a nearer parts of the scene compared to the healthy subjects (Motnikar et al., 2020). Moreover, visuospatial skills are among the best predictors of driving ability (Carr et al., 2019) and TTF relation with driving fitness is expected. Therefore, the time-to-fixate parameter is expectedly smaller in the fit-to-drive group, as it would be reasonable that they explore both close and distant zones. The absent difference between conditionally-fit and unfit-to-drive groups could be explained by a possible similar attention deficit that is common in many neurological disorders and causes patients to neglect parts of the scene (Motnikar et al., 2020). Indeed, cognitive distraction leads to delayed visual detection even in healthy samples (D'Addario & Donmez, 2019).

As expected, a greater likelihood of simulator crashes was found in previous studies (Motnikar et al., 2020; Thompson et al., 2018) in the patient population as well as a poorer driving performance. In our study, only one subject had an actual crash. An "almost-crash" situation showed expected tendency: fit- and conditionally-fit- had relatively low TTC and much less time to perform the avoidance maneuver, while this was not the case with unfit-to-drive patient. It turns out that TTC influenced the management of hazardous situations. Although TTC parameters were in the range of those in previous studies (Jurecki & Stańczyk, 2018; TTC was in the range of 0.6 s to 3 s), statistical tests did not find any significant influence of TTC, and neither IGD, nor speed to the TTF in our sample. There are two possible reasons for such a finding. Firstly, TTC and speed did not manifest larger discrepancies among the three groups of patients (Table 3). Secondly, this may be related to the cognitive processes that are responsible for TTF duration. We can conclude a similar occurrence for IGD influence on TTF and assume that TTF may be likely affected by cognitive impairment. The only statistically significant multiple interaction is among fitness, IGD, and TTC on TTF. TTC to TTF in different fitness groups may be related, as shorter TTC calls for more urgent/faster action. The logical assumption would be that TTC affects reaction (steering and braking), but we may not exclude the possible effect on TTF, especially in cases when IGD is relatively short.

Our results clearly indicate discernment among fit-to-drive patients and other groups that should be further reproduced and potentially replicated in a larger sample. It would be interesting if TTF sensitivity could be used for capturing early driving ability degeneration and risk behavior in drivers as proposed in Carr et al., (2019) or for assessment of visual attention in other complex environments. Previous report on PRTs in the same group of patients showed that patients with shorter PRT also had shorter reaction times within the standard alertness test, and at the same time PRTs weakly correlated with other neurophysiological measures leading to a conclusion that those tests are compliant to some extent (Cizman Staba et al., 2020). It was hypothesized that the failure to find stronger relations may be the result of assessment of different cognitive processes (Cizman Staba et al., 2020). This may be as well our case when PRTs are compared with the TTF – although they are related mainly for fit-to-drive patients that have both TTF and PRT low (Fig. 5a), they as well correspond to different measurement paradigms. PRTs presented in the study corresponded to the previously published results in Motnikar et al., (2020), although slight differences could be expected, as we used data from slightly different groups of patients in which we were able to obtain TTF. Overall, the PRT parameter does not depend on YOLO detector and eye-tracker data, and as such allowed us to formulate appropriate association relation with TTF that should be re-confirmed.

### Limitations of the study

This study is not without drawbacks. We recognize the following limitations in our single-parameter approach for fitness to drive assessment:

1. The simulation design could incorporate the ability for a researcher to specify time to arrival as a trigger to control the event occurrence as proposed in Chrysler et al. (2015).
2. Simulator-based driving cannot compensate for some individual preferences such as to the subjects' own vehicles to avoid adaptation to the experimental conditions (Lerner, 1993). However, our subjects did not have expectations of an exact emergency situation (collision with children), so it may be reasonable to assume that excessive adaptation to the simulator driving did not take place. This limitation could have large implications in overall method usability, as the TTF parameter cannot be repeated for many times in a manner required for rigorous eye-tracking measurement and analysis in a controlled laboratory setting (Orquin & Holmqvist, 2018).

We did not consider different crashing patterns with different variables, factors, and considerations as proposed in Chrysler et al., (2015) and Ciceri et al., 2013 for calculating and comparing TTF in neurological patients. Moreover, we did not test the pedestrian intrusion direction, as this could influence PRT duration (Jurecki & Stańczyk, 2018). Further variations for TTF calculation may as well include angles at which pedestrians appear, nighttime conditions, etc. (D'Addario & Donmez, 2019). Yet, our single scenario guaranteed that adaptation and expectation did not take place, which could be a drawback for TTF evaluation. At the same time, this is also a general limitation, as excessive repeatability required for strict scientific approach (Godwin et al., 2021) cannot be performed. However, there is still place for improvements, especially as we could not manipulate drivers' expectations and simulator scenario timing by adding arrival triggers in relation to pedestrian hazards as stated in the first limitation (Chrysler et al., 2015) or even manipulating gaze direction (Chrysler et al., 2015), which would provide a more controlled protocol (Orquin & Holmqvist, 2018). This could and should be done only to some extent, as it is of great importance to promote natural behavior in a driving simulator that cannot contain an excessive number of hazards, as in such case the drivers could be anticipating the hazard and therefore influence the TTF (Chrysler et al., 2015, Fisher et al., 2007).

3. Our sample included a heterogeneous group of neurological patients (Cizman Staba et al., 2020) that may have caused the absence of the significant differentiation between unfit- and conditionally-fit-to-drive patients. Previous studies went even further by proving the differences in response time tests between stroke patients with left and right lesions (Kaizer et al., 1988), so TTF in neurological disorders should be taken with precaution, as more thorough clinical research is required.
4. Looking at pedestrians does not necessarily mean seeing them (D'Addario & Donmez, 2019). We assumed that the gaze direction is associated with the perception of hazard onset, but this may not be the case, as both looking and not looking does not necessarily mean that something is or is not detected (Godwin et al., 2021). Moreover, it has been shown that there is a difference in abrupt and gradual hazard simulations (D'Addario & Donmez, 2019), which also affects the way subjects detect hazards.
5. The scene design should take into account possible emotional distress, especially in patients prone to such reactions, and especially in studies with multiple hazards. The subject with an actual crash in our study did not report emotional reaction to the crash incidence, but our model did not visually display the physics of a true crash once it happened, as in Chrysler et al., (2015).

## Conclusions

With simple technology such as eye-tracker video and open-source fast YOLO object detector, we were able to exploit a time-to-fixate subcomponent of perception response time in a controlled environment enabled by a driving simulator for driving performance assessment in neurological patients. The time-to-fixate parameter alone proved significantly efficient in fit-to-drive identification. For discerning among conditionally-fit and unfit-to-drive patients and for ensuring possible proposed method reproducibility, a more comprehensive approach with additional neurophysiological and driving simulator markers is needed.

**Acknowledgements** J.S. kindly acknowledges University Rehabilitation Institute Soča employees and the Nervtech team. The authors gratefully appreciate the support from Nenad B. Popović, PhD, from the University of Belgrade – School of Electrical Engineering for his valuable assistance in design of illustrations and for providing feedback on the initial manuscript structure. Also, both authors thank Nebojša Jovanović, MSc, from the University of Belgrade – School of Electrical Engineering for his kind contribution to earlier stages of the project, especially for his work on developing Python code to capture the time-to-fixate parameter. Last but not least, we are very thankful to Damjan Krstajić, founder and director of the Research Centre for Cheminformatics for his precious advice on statistical analysis in a retrospective study and to student Gregor Kovač from Faculty of Electrical Engineering, University of Ljubljana for his diligent work on YOLO application in driving simulation.

**Funding** N.M. acknowledges amiable support from the Grant No. 451-03-47/2023-01/200103 funded by the Ministry of Science, Technological Development and Innovation of the Republic of Serbia. This research was financially supported by Slovenian Research Agency within the research program ICT4QoL - Information and Communications Technologies for Quality of Life, grant number P2-0246, and the research project Neurophysiological and Cognitive Profiling of Driving Skills, grant number L2-8178.

**Data availability** Python source code and sample eye-tracker video that support the findings presented in this paper are available on GitHub under GNU General Public License v3.0 and released on Zenodo with DOI (Miljković & Sodnik, 2023a) - complete dataset of eye-tracker videos for all patients is not publicly available due to the sensitivity, but complete dataset is available from authors on reasonable request, while .csv table with relevant parameters for the complete dataset and R code for statistical analysis that support findings presented in this paper are available on Zenodo repository under Creative Commons Attribution 4.0 International (Miljković & Sodnik, 2023b).

## Declarations

were made. The images or other third party material in this article are included in the article's Creative Commons licence, unless indicated otherwise in a credit line to the material. If material is not included in the article's Creative Commons licence and your intended use is not permitted by statutory regulation or exceeds the permitted use, you will need to obtain permission directly from the copyright holder. To view a copy of this licence, visit http://creativecommons.org/licenses/by/4.0/.